\documentclass[final,1p,times]{elsarticle} 
\usepackage{graphicx} 
\usepackage{amssymb} 
\usepackage{amsthm} 
\usepackage{lineno} 

\usepackage{subfigure}

\journal{Nuclear Physics A} 
\begin{document} 

\begin{frontmatter} 


\title{Dileptons from the strongly-interacting Quark-Gluon Plasma within the
Parton-Hadron-String-Dynamics (PHSD) approach}

\author{O.~Linnyk$^{a}$, E.~L.~Bratkovskaya$^{a}$, W.~Cassing$^{b}$}

\address[a]{Institut f\"ur Theoretische Physik, %
 Johann Wolfgang Goethe University, %
 \\
 Max-von-Laue-Str. 1, %
 60438 Frankfurt am Main, %
 Germany }

\address[b]{Institut f\"ur Theoretische Physik, %
  Universit\"at Giessen, %
  \\
  Heinrich--Buff--Ring 16, %
  35392 Giessen, %
  Germany}

\begin{abstract} 
Dilepton production in $In+In$ collisions at 158 A$\cdot$GeV  is
studied within the microscopic Parton-Hadron-Strings Dynamics (PHSD)
transport approach, which is based on a dynamical quasiparticle
model (DQPM) matched to reproduce lattice QCD results in
thermodynamic equilibrium. A comparison to the data of the  NA60
Collaboration shows that the low mass dilepton spectra are well
described by including a collisional broadening of vector mesons,
while the spectra in the intermediate mass range are dominated by
off-shell
quark-antiquark annihilation in the nonperturbative QGP. In
particular, the observed softening of the $m_T$ spectra at
intermediate masses is reproduced.
\end{abstract} 

\end{frontmatter} 

\section{Introduction} \label{Intro}

Dileptons are powerful probes that deliver multi-faced
information: from the in-medium properties of hadrons to the nature
of the deconfinement phase transition and to the properties of the
deconfined state itself.
Already in 1978, E.~Shuryak proposed to use dileptons as probes of
the QGP \cite{Shuryak78} because the temperature of the plasma should
be given by the inverse slope of the expected exponential
spectral shape of its dilepton radiation. On the other hand,
dileptons are emitted over the entire space-time evolution of the
heavy-ion collision, from the initial nucleon-nucleon collisions
through the hot and dense phase and to the hadron decays after
freeze-out. This is both a challenge and advantage of the
probe.

Early concepts of the QGP were guided by the idea of a weakly
interacting system of partons which might be described by
perturbative QCD (pQCD). However, experimental observations at the
Relativistic-Heavy-Ion Collider (RHIC) indicated that the new medium
created in ultrarelativistic Au+Au collisions was interacting more
strongly than hadronic matter  and consequently this concept had to
be given up. In the present work, the dynamical evolution of the
system is described by the PHSD transport
approach~\cite{CasBrat} incorporating the off-shell propagation of
the partonic quasi-particles according to~\cite{Juchem} as well as
the transition to resonant hadronic states (or strings) in line with
Lorentz-invariant off-shell transition rates (cf~\cite{CasBrat}) for
the fusion of quark-antiquark pairs to mesonic states or three
quarks (antiquarks) to baryonic states. Dilepton radiation by
partonic off-shell quasi-particles is calculated in an effective
field theory~\cite{Linnyk06,OlenaBjorken}.
By comparing our results to experimental data, we aim to deduce the
in-medium vector meson properties as well as the properties of the
nonperturbative QGP.

\vspace{-4pt}
\section*{PHSD and dilepton emission in reactions with off-shell quarks}
\label{PHSD} \label{offshell} \vspace{-4pt}

A consistent dynamical approach -- valid also for strongly
interacting systems -- can be formulated on the basis of the
Kadanoff-Baym equations \cite{KBaym,Sascha1} or off-shell transport
equations in phase-space representation, respectively
\cite{Juchem,Sascha1}. In the Kadanoff-Baym theory the field quanta
are described in terms of propagators with complex selfenergies.
Whereas the real part of the selfenergies can be related to
mean-field potentials, the imaginary parts  provide information
about the lifetime and/or reaction rates of time-like 'particles'
\cite{Andre}. Once the proper (complex) selfenergies of the degrees
of freedom are known, the time evolution of the system is fully
governed  by off-shell transport equations (as described in Refs.
\cite{Juchem,Sascha1}).
\begin{figure}
\begin{center}
\resizebox{0.9\textwidth}{!}{%
 \includegraphics{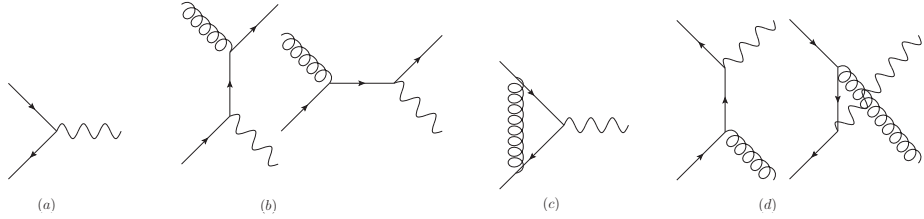}
} \caption{Diagrams contributing to the dilepton production from a
QGP: (a) Drell-Yan mechanism, (b) gluon Compton scattering (GCS),
(c) vertex correction, (d) gluon Bremsstrahlung (NLODY). Virtual
photons (wavy lines) split into lepton pairs, spiral lines denote
gluons, arrows denote quarks. In each diagram, the time runs from
left to right.} \label{diagrams}
\end{center}
\vspace{-0.5cm}
\end{figure}

The PHSD approach is a microscopic covariant transport model that
incorporates effective partonic as well as hadronic degrees of
freedom and involves a dynamical description of the hadroni\-zation
process from partonic to hadronic matter \cite{CasBrat}. Whereas the
hadronic part is essentially equivalent to the conventional HSD
approach \cite{HSD} the partonic dynamics is based on the Dynamical
QuasiParticle Model (DQPM) \cite{Cassing06} which
describes QCD properties in terms of single-particle Green's
functions (in the sense of a two-particle irreducible (2PI)
approach) and leads to effective strongly interacting partonic
quasiparticles with broad spectral functions as degrees of freedom.
The off-shell parton dynamics also allows for a solution
of the hadronization problem: the hadronization occurs by
quark-antiquark fusion or 3 quark/3 antiquark recombination which is
described by  covariant transition rates. Since the dynamical quarks
become very massive, the formed resonant 'pre-hadronic' color-dipole
states ($q\bar{q}$ or $qqq$) are of high invariant mass, too, and
sequentially decay to the ground state meson and baryon octets
increasing the total entropy. This solves the entropy problem in
hadronization in a natural way~\cite{CassXing}.
\begin{figure}
\begin{center}
\resizebox{0.75\textwidth}{!}{%
       \includegraphics{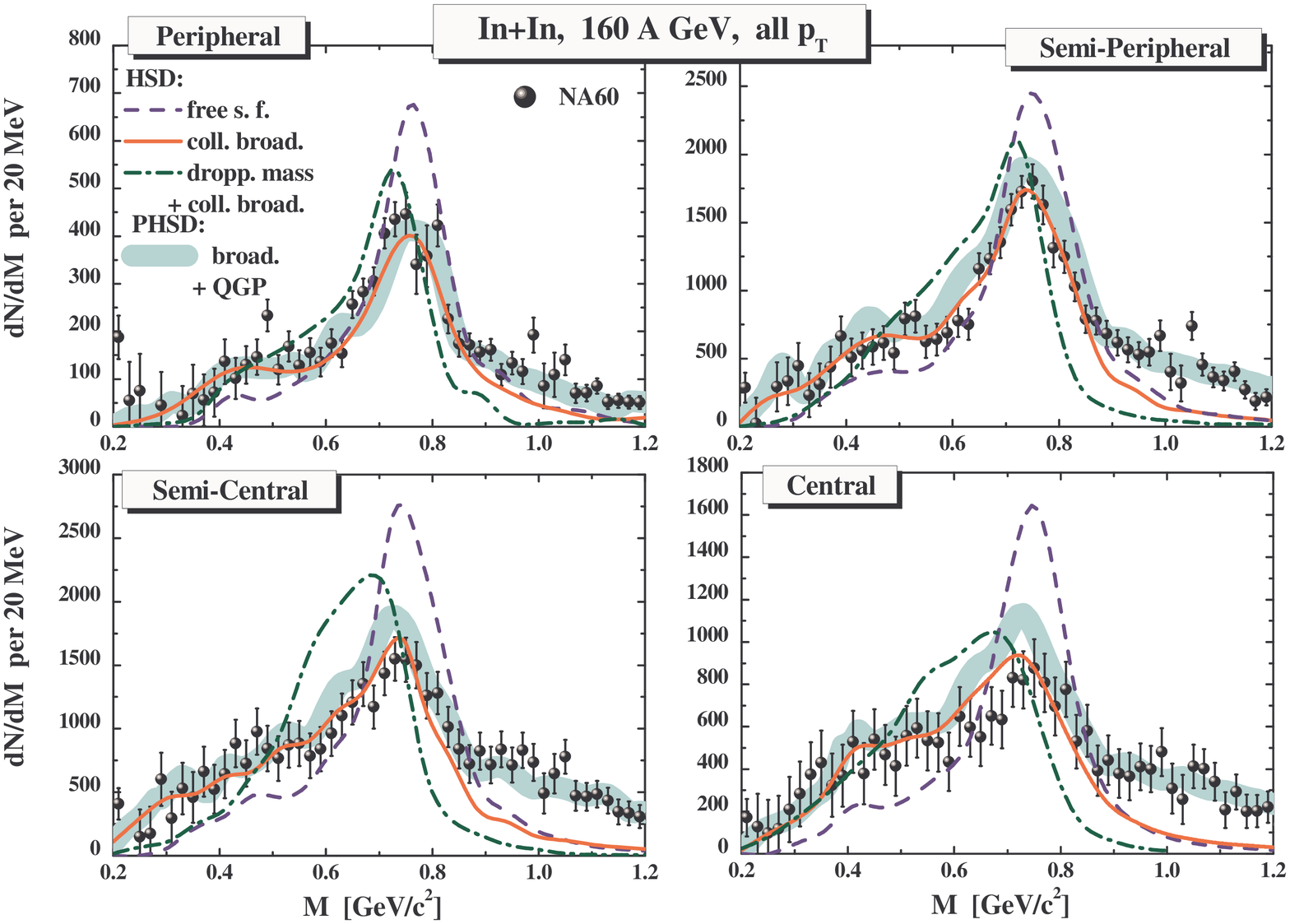}
} \caption{The HSD results for the mass differential dilepton
spectra from $In + In$ collisions at 158 A$\cdot$GeV
in comparison to the excess mass spectrum from NA60
\protect\cite{NA60}. The actual NA60 acceptance filter and mass
resolution have been incorporated \cite{Sanja}.  The solid lines
show the HSD results for a scenario including the collisional
bradening of the $\rho$-meson whereas the dashed lines correspond to
calculations with  'free' $\rho$ spectral functions for reference.
The dash-dotted lines represent the HSD calculations for the
'dropping mass + collisional broadening' model. The bands represent
the preliminary PHSD results incorporating direct dilepton radiation
from the QGP in addition to a broadened $\rho$-meson. } \label{ExcessSpectra}
\end{center}
\vspace{-0.6cm}
\end{figure}

The elementary processes involved in the dilepton radiation by the
strongly interacting QGP are illustrated in Fig.~\ref{diagrams}. The
diagrams look like those in pQCD, however, we use an effective
theory, where important modifications to pQCD are incorporated: 1) the
non-perturbative spectral functions and self-energies of the quarks,
anti-quark and gluons are taken into account ({\it i.e.} the quark
and gluon lines are dressed), 2) the running DPQM coupling
$\alpha_S$ depends on the local energy density
$\epsilon$~\cite{Cassing06} related to a temperature $T$ by the
lQCD equation of state. In particular, a non-zero width of quarks
leads to higher-twist corrections to the standard pQCD
approach~\cite{Linnyk06,OlenaBjorken}.

Note that the processes sub-leading in $\alpha_S$ are not small in
this phenomenological model. To make quantitative comparison to
data, we include the following $O(\alpha_S)$ partonic processes as
sources of the dileptons in addition to the simple leading order
Drell-Yan $q+\bar q$ annihilation mechanism: Gluon Compton
scattering ($q+g\to \gamma^*+q$ and $\bar q+g\to \gamma^*+\bar q$)
and quark + anti-quark annihilation with gluon Bremsstrahlung in the final
state ($q+\bar q\to g+\gamma^*$). By implementation of the off-shell
cross sections into the PHSD transport we can calculate the
dilepton radiation from the nonequilibrium strongly-interacting QGP.

\vspace{-4pt}
\section*{Comparison to NA60 data} \label{rho} \label{data}
\vspace{-4pt}

Let us start with results on the
in-medium properties of vector mesons and move to the dilepton
radiation from the QGP later on. Various models predict that
hadrons change in the (hot and dense) nuclear medium;
in particular, a broadening of
the spectral function or a mass shift of the vector mesons was
expected to be considerable. Furthermore, QCD sum rules indicated that a mass
shift may lead to a broadening and vice versa~\cite{MuellerSumRules};
therefore  both modifications  should
be studied simultaneously, too. Thus we explore three possible scenarios:
 (1) a broadening of
the $\rho$ spectral function, (2) a mass shift, and (3) a broadening plus a
mass shift. The HSD off-shell transport approach allows to investigate in a
consistent way the different scenarios for the modification of
vector mesons in a hot and dense medium. In the off-shell transport,
meson spectral functions change dynamically during the propagation
through the medium and evolve towards the on-shell spectral function
in the vacuum.

As we find in Fig. \ref{ExcessSpectra} the NA60 data favor the
scenario of the in-medium broadening of vector mesons.  Note that in
the data - presented in this
plot - the D-meson contribution has not been subtracted. The NA60
collaboration has published acceptance corrected data with
subtracted charm contribution recently. The comparison to the new
data set leads to the same conclusions as here and will be presented
in an extended publication \cite{YYY}. Note that a comparison of
HSD results for the free case and the three in-medium scenarios to
the CERES data in~\cite{OurDilept08} also has shown that the
spectrum is described better, if a broadening of the $\rho$-meson
spectral function in the medium is assumed ({cf.} Fig. 4
in~\cite{OurDilept08}).
\begin{figure*}
\begin{center}
\subfigure
{
    \resizebox{0.54\textwidth}{!}{%
       \includegraphics{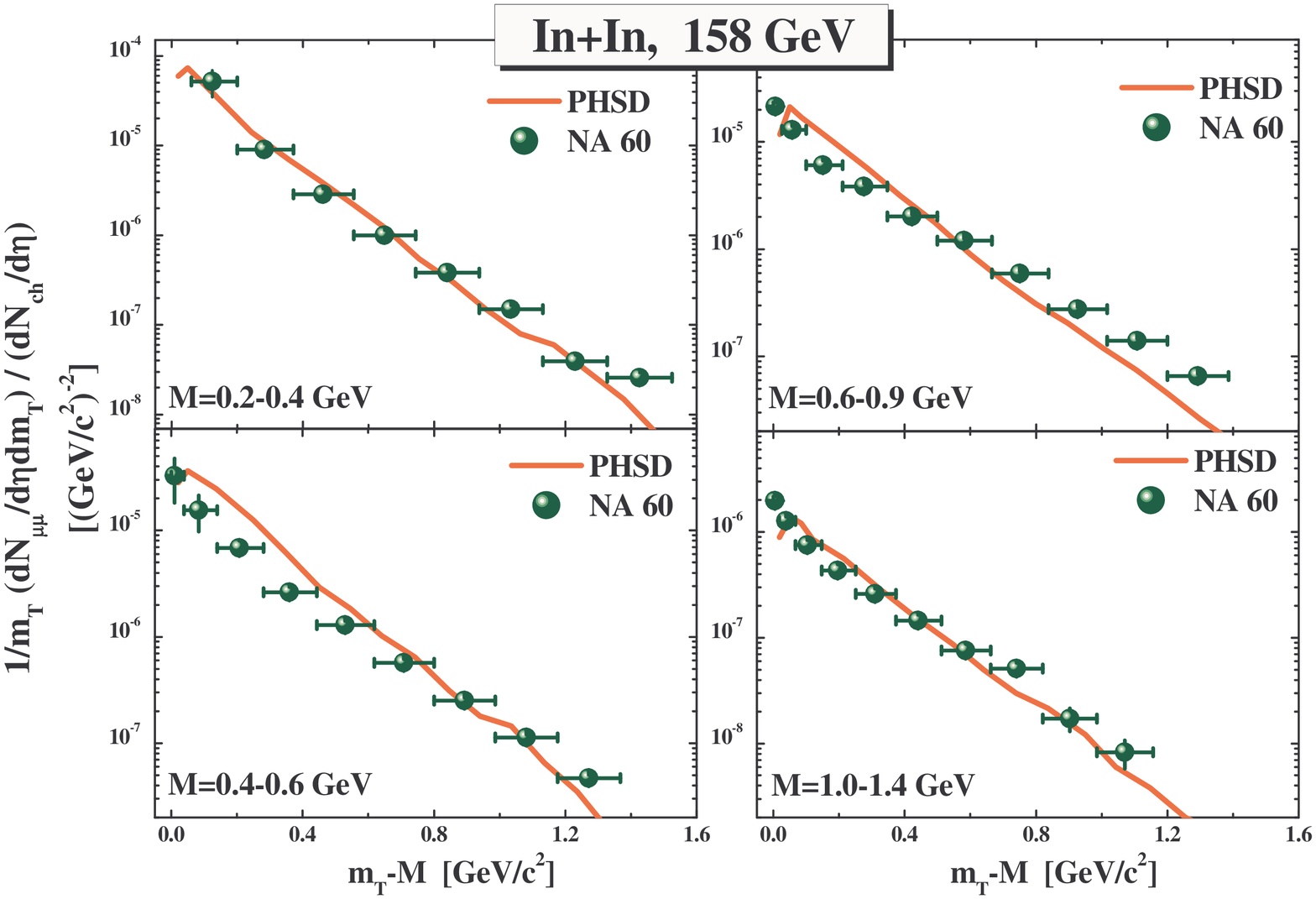}
    }
\label{a} } \hspace{-0.3 cm} 
\subfigure
{
   \resizebox{0.44\textwidth}{!}{%
        \includegraphics{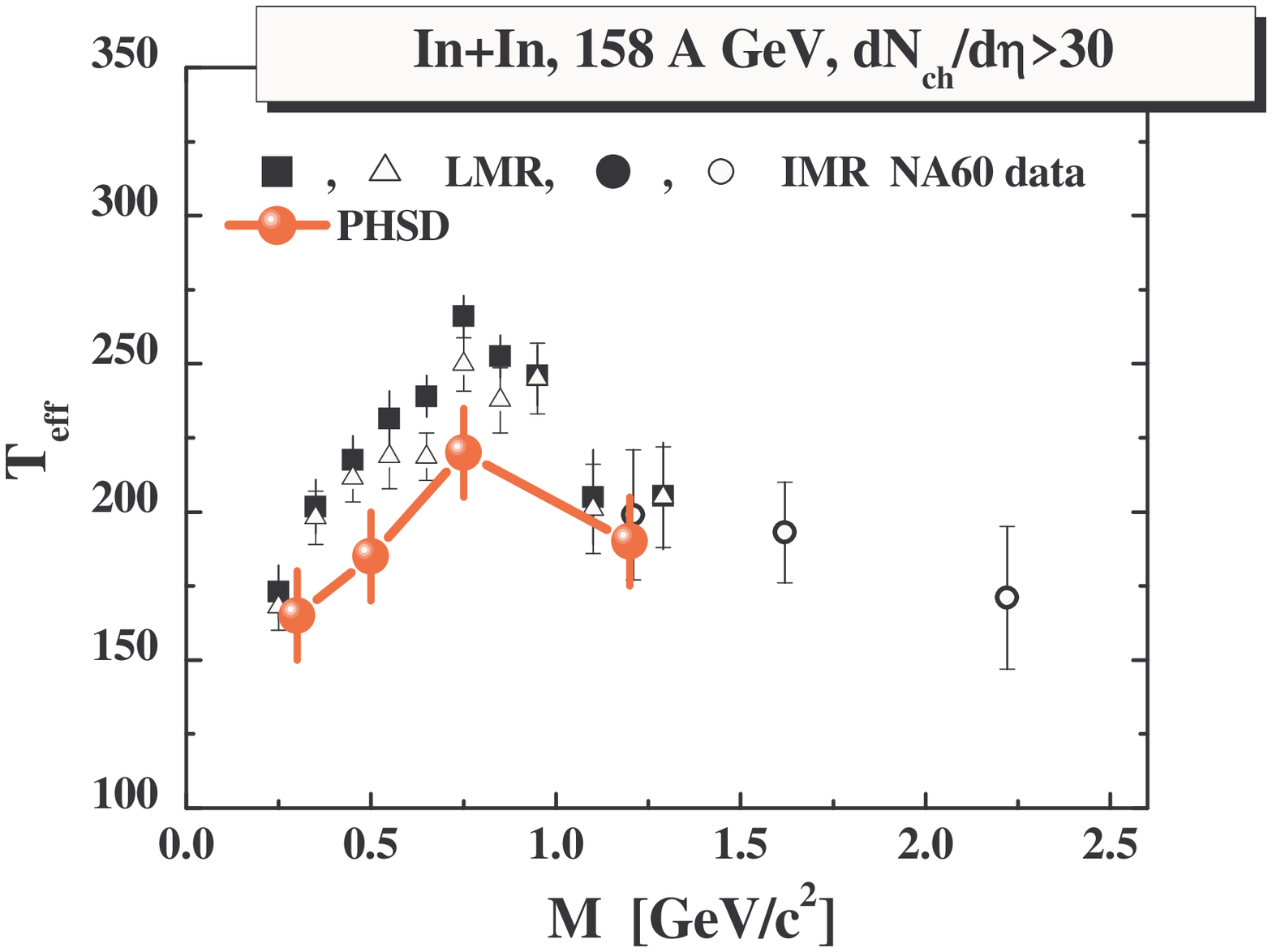}
   } \label{b}
} \caption{Left: Transverse mass spectra of dileptons for In+In at 158 A$\cdot$GeV. Right:
Evolution of the inverse slope parameter $T_{eff}$ of the dimuon
yield with invariant mass.} \label{Slopes}
\end{center}
\vspace{-0.6cm}
\end{figure*}

On the other hand, there is a discrepancy between the HSD (hadronic)
results and the data in the intermediate of mass range above 1~GeV. This
discrepancy is not accounted for by hadronic sources in HSD  --
in-medium or free -- and might be seen as a signal of partonic
matter, manifest already at the top SPS energy. The calculation of the
QGP contribution in PHSD  fully accounts for the observed
excess (cf. the bands in Fig.~\ref{ExcessSpectra}).

The intriguing finding of the NA60 Collaboration is that the
effective temperature of the dileptons in the intermediate  mass
range is lower than the $T_{eff}$ of the dileptons from the hadronic
phase! This implies that the quark-antiquark annihilation (or
partonic channels) occur dominantly before the collective radial
flow has developed. This feature of the data is also reproduced in
PHSD (cf. Fig.~\ref{Slopes}, rhs). A detailed look at the PHSD
results confirms that the lower mass region is dominated by hadronic
sources while the spectrum in the intermediate mass range is
dominated by the off-shell partonic channels in the QGP~\cite{YYY}.

\vspace{-8pt}
\section*{Acknowledgements}
\vspace{-4pt} \hspace{-18pt} OL acknowledges
support within the ``HIC for FAIR" framework of the ``LOEWE"
program. \vspace{-8pt}



\begin{thebibliography}{00} 
\bibitem{Shuryak78}
    E.V. Shuryak, Phys. Lett. B78 (1978)  150.
\bibitem{CasBrat} W. Cassing and E. L. Bratkovskaya,
    {\em Phys. Rev.} C 78 (2008) 034919.
\bibitem{Juchem}
      W. Cassing and S. Juchem, {\em Nucl. Phys.} A 665 (2000) 377;
      {\it ibid} A 672 (2000) 417.
\bibitem{Linnyk06}
  O.~Linnyk, S.~Leupold and U.~Mosel,
  Phys.\ Rev.\  D { 75} (2007)  014016.
\bibitem{OlenaBjorken}
      O.~Linnyk,{\em in preparation}
\bibitem{KBaym}
      L. P. Kadanoff, G. Baym, {\it Quantum Statistical Mechanics},
      Benjamin, 1962.
\bibitem{Sascha1}
      W. Cassing and S. Juchem, Nucl. Phys. A 672 (2000) 417; S. Juchem {\it et
      al.}, Nucl. Phys. A {743} {2004} 92.
\bibitem{Andre}
      A. Peshier and W. Cassing, { Phys. Rev. Lett.} 94 (2005) 172301.
\bibitem{HSD}
      W. Cassing and E. L. Bratkovskaya, { Phys. Rept.} 308 (1999) 65;
      W. Ehehalt and W. Cassing, {Nucl. Phys.} A 602 (1996) 449.
\bibitem{Cassing06}
      W. Cassing, {Nucl. Phys.} A 791 (2007) 365; {\it ibid.}  A 795 (2007) 70.
\bibitem{CassXing}
 W.~Cassing, E.~L.~Bratkovskaya and Y.~Xing, Prog. Part. Nucl.
 Phys. 62 (2009)  359.
\bibitem{MuellerSumRules}
  J.~Ruppert, T.~Renk and B.~M{\"u}ller,
  Phys.\ Rev.\  C {73} (2006)  034907.
\bibitem{NA60}
      R. Arnaldi {\it et al.}, NA60 Collaboration,
              Phys. Rev. Lett. {96} (2006)   162302;
      J. Seixas {\it et al.},  J. Phys. G { 34} (2007) S1023;
       S. Damjanovic {\it et al.}, Nucl. Phys. { A 783} (2007) 327c.
\bibitem{Sanja}
      S. Damjanovic, private communication.
\bibitem{YYY}
      O.~Linnyk, E.~Bratkovskaya, W.~Cassing, {\it in preparation}
\bibitem{OurDilept08}
  E.~L.~Bratkovskaya, W.~Cassing and O.~Linnyk,
  Phys.\ Lett.\  B { 670} (2009)  428.
\end{thebibliography}
\end{document}